# A Dynamical Effective Medium Theory for Elastic Metamaterials

Ying Wu, Yun Lai, and Zhao-Qing Zhang[*]

Department of Physics, Hong Kong University of Science and Technology, Clear Water Bay, Kowloon, Hong Kong, China

Abstract

We develop a dynamical effective medium theory to accurately predict the unusual properties of elastic metamaterials in two dimensions near the resonant frequencies. The theory shows that the effective bulk modulus $\kappa_e$, shear modulus $\mu_e$, and mass density $\rho_e$ can be made negative by choosing proper resonant scatterers, leading to 8 possible types of wave propagation. The theory not only provides a convenient tool to search for various metamaterials with desired properties, but also gives a unified physical picture of these properties. Here we demonstrate two examples. One possesses large band gaps at low frequencies. The other exhibits two regions of negative refraction: one for both longitudinal and transverse waves and the other for longitudinal waves only.

---

[*] To whom correspondence should be addressed. Email:phzzhang@ust.hk

**I Introduction**

During the past decade, the propagation of classical waves in complex media has seen enormous developments. Recently, the artificial subwavelength electromagnetic (EM) materials, denoted metamaterials, have significantly broadened the range of material responses found in nature.[1,2] Through the use of different metamaterials, various novel concepts have been successfully demonstrated, such as left-handed (LH) materials, which exhibit simultaneous negative permittivity and permeability.[1,2] LH materials can support unusual phenomena like negative refraction[1] and sub-wavelength focusing.[2] The working principle behind these unusual phenomena is the concept of homogenization. It has been successfully demonstrated that the double-negative metamaterial behaves like a homogeneous medium capable of producing negative effective permittivity, $\varepsilon_{eff}$, and effective permeability, $\mu_{eff}$, simultaneously in a frequency range above the resonance of the building blocks of the metamaterial.[3] Knowing these effective parameters is also useful in studying wave propagation through a finite-sized metamaterial. For example, in the study of perfect lens, it is simple and convenient to represent a slab of metamaterial by an effective medium with $\varepsilon_{eff}$ and $\mu_{eff}$ in the Maxwell equation.[2] Since the unusual properties of a metamaterial are derived from the built-in resonances in its constituent units, an effective medium theory is required to be valid near resonances in order to describe the metamaterial properly. When the constituent units are made of isotropic scatterers, such a dynamical effective medium (DEMT) theory has been known for long time.[4] Recently, the acoustic analogue of metamaterials has also been explored.

An acoustic metamaterial made of silicon rubber spheres suspended in water was demonstrated to possess simultaneous negative effective mass density and modulus.[5]

Although various DEMT have been proposed for EM[4][6][7][8] and acoustic[5][9] metamaterials, much less is studied for elastic metamaterials. A kind of locally resonant phononic crystal which can support large band gaps at low frequencies has been reported recently.[10][11] By "elastic metamaterials", we mean that the host material is a solid instead of a fluid, which makes the whole system elastic for small stresses. The existence of three independent physical parameters in isotropic solids can lead to three effective parameters, i.e., effective bulk modulus, $\kappa_e$, effective shear modulus, $\mu_e$, and effective mass density, $\rho_e$, giving rise to much richer physics. As a general property of metamaterials, we expect the elastic metamaterials can have negative effective parameters, which are outside both the traditional bound determined by values of the constituents and the more sophisticated Hashin and Shtrikman bounds. The negativity of each single effective parameter and their various combinations can produce eight types of wave propagation, as will be discussed later. A DEMT in this case can provide a unified physical understanding of eight types of elastic metamaterials under the same framework. However, to date, only a few types of elastic metamaterials were explored due to the lack of information on the resonant properties of the constituent unit which determines the effective medium parameters. An elastic DEMT, which directly relates the resonances of the constituent unit to the effective medium parameter, can, therefore, also provide a useful guidance to the search of metamaterial with any desired properties of wave propagation. The purpose of this work is to develop a DEMT for elastic metamaterials in two dimensions.

In this work, we present a DEMT for certain two-dimensional elastic metamaterials consisting of cylindrical scatterers arranged in a triangular lattice embedded in a solid host. The theory directly relates the effective medium parameters, $\kappa_e$, $\mu_e$ and $\rho_e$, to the scattering properties of the scatterers. In order to check the validity of our theory, we have also performed the band-structure calculations of the metamaterials. Our results show that the theory is capable of accurately reproducing the band structure results in the long wavelength limit, even in the vicinities of the built-in resonances. Furthermore, we show that our theory can also provide a useful guidance to search for various metamaterials of desired properties by altering the scattering properties of the scatterers. Here we demonstrate the design of two types of metamaterials: one possessing large band gaps at low frequencies and the other exhibiting two regions of negative refraction.

The derivation of the DEMT is presented in Section II. The application of the DEMT to a simple metamaterial is given in Section III. Section IV considers two more complicated metamaterials, in which simple scatterers are replaced by coated scatterers. Finally, a conclusion is given in Section V.

## II  Effective Medium Theory

We consider a 2D elastic metamaterial composed of cylindrical inclusions of radius $r_s$ with $(\rho_s, \mu_s, \kappa_s)$ imbedded in a homogeneous matrix of $(\rho_0, \mu_0, \kappa_0)$. Here $\rho$ denotes mass density, $\mu$ represents shear modulus and $\kappa$ denotes bulk modulus satisfying $\kappa = \lambda + \mu$, where $\lambda$ represents the Lamé constant. Due to the translational symmetry along the cylinder's axis, say z-axis, the elastic modes in the system can be decoupled into the z-mode and the xy-mode, with vibrations along the z-axis and in the x-y plane, respectively. Here, we consider the more complicated xy-mode. The displacements $\vec{u}$ in x-y plane can be described by the following wave equation,[12]

$$\rho(\vec{r})\frac{\partial u_i^2(\vec{r})}{\partial t^2} = \nabla \cdot (\mu(\vec{r})\nabla u_i(\vec{r})) + \nabla \cdot \left(\mu(\vec{r})\frac{\partial \vec{u}(\vec{r})}{\partial x_i}\right) + \frac{\partial}{\partial x_i}[\lambda(\vec{r})\nabla \cdot \vec{u}(\vec{r})], \qquad (1)$$

where $\vec{u}$ is the displacement field. In general, $\vec{u}$ can be decoupled into longitudinal part and transverse part, i.e. $\vec{u} = \nabla \phi_l + \nabla \times (\phi_t \hat{e}_z)$, where $\phi_l$ and $\phi_t$ are the longitudinal and transverse gauge potentials, respectively. In the frequency domain, the solutions to $\phi_l$ and $\phi_t$ can be written as $\phi_\alpha(s) = \sum_m a_{\alpha m}(s) J_m(k_{\alpha s} r) e^{im\theta}$  $(\alpha = l, t)$ in the cylinder and $\phi_\alpha(0) = \sum_m a_{\alpha m}(0) J_m(k_{\alpha 0} r) e^{im\theta} + b_{\alpha m}(0) H_m^{(1)}(k_{\alpha 0} r) e^{im\theta}$, $(\alpha = l, t)$ in the matrix. Here $J_m(x)$ and $H_m^{(1)}(x)$ are, respectively, the Bessel and Hankel functions of the first kind, and $k_{l\delta} = \omega\sqrt{\rho_\delta/(\lambda_\delta + 2\mu_\delta)} = \omega\sqrt{\rho_\delta/(\kappa_\delta + \mu_\delta)}$, $k_{t\delta} = \omega\sqrt{\rho_\delta/\mu_\delta}$ are, respectively, the longitudinal and transverse wave vectors in the cylinder $(\delta = s)$ or in the matrix $(\delta = 0)$, and $\omega$ is the angular frequency. The longitudinal and transverse waves in the matrix are coupled by the scatterings of the embedded cylinders. To determine the coefficients of those Bessel and Hankel functions, we need to consider the elastic boundary conditions which are the continuities of the radial and tangential component of the displacement field, i.e. $u_r$ and $u_\theta$, plus the continuities of the stresses, $\sigma_{rr}$ and $\sigma_{r\theta}$, at the interface. The stresses are defined as

$$\sigma_{rr} = (\lambda + 2\mu)\frac{\partial u_r}{\partial r} + \lambda(\frac{1}{r}\frac{\partial u_\theta}{\partial \theta} + \frac{u_r}{r}) \quad \text{and} \quad \sigma_{r\theta} = \mu(\frac{1}{r}\frac{\partial u_r}{\partial \theta} + \frac{\partial u_\theta}{\partial r} - \frac{u_\theta}{r}) \quad .$$

These continuities on the surface of a cylinder relate $b_{lm}(0)$ and $b_{tm}(0)$ to $a_{lm}(0)$ and $a_{tm}(0)$ through $b_{lm}(0) = D_m^{ll}(s,0)a_{lm}(0) + D_m^{lt}(s,0)a_{tm}(0)$ and $b_{tm}(0) = D_m^{tl}(s,0)a_{lm}(0) + D_m^{tt}(s,0)a_{tm}(0)$, where $D_m^{\alpha\beta}(s,0)$ $(\alpha,\beta = l,t)$ are the Mie-like scattering coefficients determined by the boundary conditions on the interface between the cylinder and matrix. $D_m^{\alpha\beta}(s,0)$ $(\alpha,\beta = l,t)$ are all functions of $k_{ls}, k_{ts}, k_{l0}, k_{t0}$ and $r_s$.[13] The explicit expressions for $D_m^{\alpha\beta}(s,0)$ are displayed in Appendix A. In the long wavelength limit, we derive the effective medium formulae for such a system by considering the scattering of elastic waves by a coated cylinder embedded in the effective medium with effective parameters $(\kappa_e, \mu_e, \rho_e)$.[5,8,14,15] Fig.1 shows the coated cylinder consisting of the cylindrical scatterer surrounded by a shell of the matrix. The inner and outer radii, which are denoted by $r_s$ and $r_0$, respectively, satisfy $\pi r_s^2 / \pi r_0^2 = p$, where $p$ is the filling fraction of the system. The effective parameters $\kappa_e$, $\mu_e$ and $\rho_e$ are determined by the condition that the total scattering of the coated cylinder vanishes. This condition together with the boundary conditions on the surface of the coated cylinder at $r = r_0$ provides another two relations: $b_{lm}(0) = D_m^{ll}(e,0)a_{lm}(0) + D_m^{lt}(e,0)a_{tm}(0)$ and $b_{tm}(0) = D_m^{tl}(e,0)a_{lm}(0) + D_m^{tt}(e,0)a_{tm}(0)$, where $D_m^{\alpha\beta}(e,0)$ ($\alpha,\beta = l,t$) can be obtained by replacing $\lambda_s$, $\rho_s$, $\mu_s$, $k_{ls}$, $k_{ts}$ and $r_s$ in $D_m^{\alpha\beta}(s,0)$ ($\alpha,\beta = l,t$) with $\lambda_e$, $\rho_e$, $\mu_e$, $k_{le}$, $k_{te}$ and $r_0$, respectively. Here $k_{le}$ and $k_{te}$ are the longitudinal and transverse wave vectors in the effective medium. By comparing these with the previous relations, we obtain $D_m^{\alpha\beta}(e,0) = D_m^{\alpha\beta}(s,0)$ ($\alpha,\beta = l,t$). In the long wavelength limit of $k_{l0}r_0 \ll 1$, $k_{t0}r_0 \ll 1$, $k_{le}r_0 \ll 1$ and $k_{te}r_0 \ll 1$,[16] we have the following effective medium equations for the elastic metamaterial:

$$\frac{(\kappa_0 - \kappa_e)}{(\mu_0 + \kappa_e)} = \frac{4\tilde{D}_0^{ll}(s,0)}{i\pi r_0^2 k_{l0}^2}, \tag{3a}$$

$$\frac{(\rho_0 - \rho_e)}{\rho_0} = -\frac{8\tilde{D}_1^{ll}(s,0)}{i\pi r_0^2 k_{l0}^2}, \tag{3b}$$

$$\frac{\mu_0(\mu_0 - \mu_e)}{(\kappa_0\mu_0 + (\kappa_0 + 2\mu_0)\mu_e)} = \frac{4\tilde{D}_2^{ll}(s,0)}{i\pi r_0^2 k_{l0}^2}, \tag{3c}$$

where $\tilde{D}_m^{ll}(s,0)$ is the $D_m^{ll}(s,0)$ after taking the leading terms of $J_m(x)$ and $H_m^{(1)}(x)$ with $x \equiv k_{l0}r_s$ and $k_{t0}r_s$. The derivation of Eq. (3) is given in Appendix B. It is interesting to point out that $\kappa_e$, $\rho_e$ and $\mu_e$ are independently determined by $\tilde{D}_m^{ll}(s,0)$ of the embedded cylinders alone with $m = 0$, 1 and 2, respectively. If we further consider the condition of $k_{ls}r_s \ll 1$, $k_{ts}r_s \ll 1$ in $\tilde{D}_m^{ll}(s,0)$, the quasi-static limit is reached and Eq.(3) becomes:

$$\frac{(\kappa_0 - \kappa_e)}{(\mu_0 + \kappa_e)} = p\frac{(\kappa_0 - \kappa_s)}{(\mu_0 + \kappa_s)}, \tag{4a}$$

$$(\rho_0 - \rho_e) = p(\rho_0 - \rho_s), \tag{4b}$$

$$\frac{(\mu_0 - \mu_e)}{(\kappa_0\mu_0 + (\kappa_0 + 2\mu_0)\mu_e)} = p\frac{(\mu_0 - \mu_s)}{(\kappa_0\mu_0 + (\kappa_0 + 2\mu_0)\mu_s)}, \tag{4c}$$

where $p$ is the filling fraction of the scatterer. The 3D results were reported by Berryman decades ago. It should be pointed out that elastic EMT cannot recover the acoustic EMT by setting all the shear moduli to be zero, because the boundary conditions of two waves are different.

In nature, no elastic material has negative mass density or negative bulk/shear modulus. However, for metamaterials, it is possible to achieve negativity in any of $\kappa_e$, $\rho_e$ and $\mu_e$, when frequency is near the resonant frequencies of $\tilde{D}_m^{ll}(s,0)$, $m = 0$, 1, 2. These negativities as well as their combinations can give rise to various interesting properties unseen in natural material. For instance, since the effective phase velocities are $c_{le} = \sqrt{\lambda_e + 2\mu_e}\sqrt{1/\rho_e} = \sqrt{\kappa_e + \mu_e}\sqrt{1/\rho_e}$ and $c_{te} = \sqrt{\mu_e}\sqrt{1/\rho_e}$ for longitudinal and transverse waves, respectively, a single negativity in $\rho_e$ in a frequency window leads to imaginary $c_{le}$ and $c_{te}$, which in turn implies band gap for both longitudinal and transverse waves. However, a double negativity in both $\rho_e$ and $\mu_e$ makes the refractive index negative for the transverse waves. In Table 1, we list 8 possible types of elastic metamaterials with different combinations of signs in $\kappa_e + \mu_e$, $\rho_e$ and $\mu_e$.

## III  Application to Simple Metamaterials

In order to search for various elastic metamaterials, we start from the following simple metamaterial, i.e. a triangular lattice of silicon rubber cylinders with radius $0.3a$ ($a$ is the lattice constant) embedded in epoxy host.[17] Since silicon rubber possesses very small longitudinal and transverse wave speeds, Mie-like resonances may occur at low frequencies, which serve as the built-in resonances required for metamaterials. In Fig. 2(a), we show the band-structures near the Γ point calculated by the multiple-scattering method[13 18] in dimensionless frequency $\tilde{f} = \frac{\omega}{2\pi} \frac{a}{c_{t0}}$, where $c_{t0} = \sqrt{\mu_0}\sqrt{1/\rho_0}$. Multi-gap structures associated with various resonances are clearly seen in the frequency range $\tilde{f} \leq 0.05$. In this system, the long wave length conditions hold when $\tilde{f} \ll 0.3$. In Fig. 2(b), (c) and (d), we plot the real parts of $\kappa_e$, $\rho_e$ and $\mu_e$ calculated from Eq. (3), respectively. The imaginary parts are 15 orders of magnitude smaller and can be ignored. It is clear that Fig. 2(b)-(d) exhibit very rich resonant behaviors for each of the three effective medium parameters even at such low frequencies. The dispersion curves of the system can also be obtained from Eq. (3) through $\omega = \left(\sqrt{(\kappa_e + \mu_e)}/\sqrt{\rho_e}\right)K$ (longitudinal) and $\omega = \left(\sqrt{\mu_e}/\sqrt{\rho_e}\right)K$ (transverse). These results are also plotted in Fig. 2(a) in solid curves. The excellent agreement between the open circles and solid (dashed) curves shows the validity and accuracy of Eq. (3). In Fig. 2(a), we also plot in dotted lines the dispersion curves obtained from the quasi-static theory, i.e. Eq.(4), which gives the following frequency independent effective parameters: $\rho_e = 1.0333\rho_0$, $\lambda_e = 0.24824\lambda_0$ and $\mu_e = 0.44935\mu_0$. Obviously, the quasi-static effective medium theory fails to predict the band-structures above the lowest resonance. It should be pointed out that the dispersion relations for elastic waves are in general not isotropic near the Γ point even at low frequencies.[19] However, it can be shown analytically that isotropy holds for the case of triangular lattice considered here. When the dispersion is anisotropic such as in a square lattice, our DEMT can only predict the angle-averaged dispersion relation at low frequency.[20] Even in this case, the gap positions can still be accurately predicted. In Fig. 2, we find a few narrow band gaps for both longitudinal and transverse waves in the regions of single negativity in $\rho_e$. Two obvious ones are at $\tilde{f}$ = (0.01871, 0.02011) and (0.02222, 0.02391). We also find regions of single negativity in $\kappa_e$ and $\mu_e$. However, we do not find any regions of double negativity. Even such a simple system can already produce various kinds of wave propagation listed in Table 1. However, the regions of negativity shown in Fig. 2 are too narrow to be useful. The next question is whether we can use Eq. (3) as a guide to engineer the scatterers to search for more interesting properties.

## IV Application to Complicated Metamaterials

In the following two examples we show how to engineer the scatterers (or $\tilde{D}_m^{ll}(s,0)$) in a way to (i) widen the gaps shown in Fig. 2(a), which is equivalent to enlarging the negative regions of $\rho_e$, and (ii) create a region of double negativity, which is equivalent to making the negative regions of both $\rho_e$ and $\mu_e$ or/and $\kappa_e + \mu_e$ overlap. For these purposes, we need to understand the mode structures near the resonances of $\rho_e$, $\mu_e$ and $\kappa_e$. In Fig. 3(a), (b) and (c), we plot the displacement fields of the Bloch states at the $\Gamma$ point corresponding to the lowest resonances of $\rho_e$, $\mu_e$ and $\kappa_e$, respectively, as marked by "A", "B" and "C" in Fig.2. The brightness and the arrows indicate, respectively, the magnitude and the direction of the displacement field $\bar{u}$ at a certain time. Fig. 3(a) shows that the core part of silicon rubber scatterer oscillates as a whole. This mode can be regarded as a simple "mass-spring" harmonic oscillator, with the core part serving as "mass" and the edge part serving as "spring". Fig. 3(b) exhibits that the shape of the silicon rubber cylinder is deformed in oscillation with its cross-sectional area conserved and Fig. 3(c) shows that the shape of the silicon rubber cylinder is fixed, with its cross-sectional area oscillating in time. The above mode structures provide useful guidance for the engineering of the silicon rubber cylinders. For instance, replacing the inner region of silicon rubber with another heavier cylinder, e.g. lead[17], will enhance the field oscillation of the silicon rubber, which in turn will widen the resonant region of $\rho_e$. It is straightforward to generalize the effective medium theory of Eq.(3) to the case where the scatterers are layered cylinders. In this case, the quantities $\tilde{D}_m^{ll}(s,0)$ can be obtained by using the standard transfer-matrix method.[13] In Fig. 4(a), we show the results of both band-structure calculation (in open circles) and Eq. (3) (in solid and dashed curves) when the radius of lead cylinder is $r_{lead} = 0.9 r_s$. Compared to Fig. 2(a), it is interesting to see that the lead replacement has significantly widened the negative region of $\rho_e$ at the first resonance as shown in Fig. 4(b) and a large band gap with gap-midgap ratio $\Delta \tilde{f} / \tilde{f}_c = 0.74$ is created. Thus, we can use Eq. (3) to search for metamaterials with large $\Delta \tilde{f} / \tilde{f}_c$. Fig. 4(c) shows the calculated $\Delta \tilde{f} / \tilde{f}_c$ as a function of $r_{lead} / r_s$ at different values of $r_s / a$. It is seen that a ratio as large as $\Delta \tilde{f} / \tilde{f}_c = 1.2$ can be achieved. It should be mentioned that the above 3-component metamaterial has been studied before[10]. Recently, the analysis of a simple model based on a stiff matrix also suggests a negative effective mass density in such materials[11]. However, the model cannot provide any information on the effective moduli.

In the following, we use Eq.(3) to search for elastic metamaterials with negative refractive index. In this case, we need to consider the resonances of both $\rho_e$ and $\mu_e$ or/and $\kappa_e$. Fig. 3(b) and (c) suggest that if we want to enhance these resonances, we need to make the scatterer easier to deform and compress. Thus, we replace the core region of the silicon rubber by air.[17] The radius of air core $r_{air}$ is chosen such that the negative regions of both $\rho_e$ and $\mu_e$ or/and $\kappa_e$ overlap. The results of band-structure calculation with $r_{air} = 0.87 r_s$ are shown in Fig. 5(a) in open circles, which coincide with the results of Eq.(3) shown in solid and dashed curves. In the region of $0.12240 < \tilde{f} < 0.12253$, negative-$n$ bands of both longitudinal and transverse waves are found, which implies negativities in both $\rho_e$ and $\kappa_e + \mu_e$ and $\mu_e$. These negativities are induced only by the resonances of $\rho_e$ and $\mu_e$ as shown in Fig. 5(b). While in another region of $0.1234 < \tilde{f} < 0.12356$, negative-$n$ band is found only for longitudinal waves, which implies double negativity in $\rho_e$ and $\kappa_e + \mu_e$, arising from resonances of both $\rho_e$ and $\kappa_e$. We notice that there exist small discrepancies between the band-structure calculation and the effective medium prediction for the longitudinal branches. The reason is that the values of $k_{l0} r_0$ and $k_{t0} r_0$ are not much less than 1 for the frequencies shown in Fig. 5(a). Thus the approximation we used to derive Eq.(3) becomes less accurate. For example, when $\tilde{f} = 0.1230$, $k_{l0}r_0 = 0.185$ and $k_{t0}r_0 = 0.406$, the errors induced by taking leading terms are 5% in $H_0^{(1)}(k_{l0}r_0)$, 0.1% in $H_1^{(1)}(k_{l0}r_0)$, 0.02% in $H_2^{(1)}(k_{l0}r_0)$, 3% in $H_0^{(1)}(k_{t0}r_0)$, 0.9% in $H_1^{(1)}(k_{t0}r_0)$ and 0.2% in $H_2^{(1)}(k_{t0}r_0)$. Thus, the errors introduced are larger in $H_0^{(1)}$ than in $H_1^{(1)}$ and $H_2^{(1)}$. In other words, $\kappa_e$ is less accurate than $\rho_e$ and $\mu_e$. Therefore, the discrepancies between band-structure calculation and effective medium prediction can only be found for the longitudinal branches, which involves $\kappa_e$, but not for transverse branches.

## V  Conclusion

In conclusion, we have developed a dynamical effective medium theory for elastic metamaterials. The effective parameters, i.e. $\kappa_e, \rho_e$ and $\mu_e$, are independently determined and can turn negative near resonances. By comparing our theory with band-structure calculations of three kinds of elastic metamaterials, we show explicitly that our theory is accurate at low frequencies even near resonances whereas the quasi-static theory fails. The theory allows us to predict various kinds of wave propagation arising from different kinds of negativity in the effective parameters. For example, negativity in $\rho_e$ only can bring band gaps for both longitudinal and transverse branches. Negativities in both $\rho_e$ and $\mu_e$ and/or $\kappa_e + \mu_e$ will introduce double negative bands for elastic waves. The theory can also provide a convenient tool to search for various metamaterials with desired properties. Here we have demonstrated two examples of designing elastic metamaterials: one has large band gap at low frequencies and the other exhibits double negative bands for both longitudinal and transverse branches. Through these two examples, we show that our theory not only provides a convenient tool to search for metamaterials of various desired interesting properties but also gives a unified picture to understand the physical origins of these properties.

The work was supported by Hong Kong RGC Grant No. 604703.

**Appendix A**: Calculating the scattering coefficients $D_m^{\alpha\beta}(s,0)$

The coefficients $D_m^{\alpha\beta}(s,0)$, determined by boundary conditions, connect the scattered field to the incident field at each scatterer. The explicit expression of $D_m^{\alpha\beta}(s,0)$ takes the following form:

$$D_m^{ll}(s,0) = \frac{1}{M_m}\begin{vmatrix} a_1^l & b_1^t & c_1^l & c_1^t \\ a_2^l & b_2^t & c_2^l & c_2^t \\ a_3^l & b_3^t & c_3^l & c_3^t \\ a_4^l & b_4^t & c_4^l & c_4^t \end{vmatrix}, \quad D_m^{lt}(s,0) = \frac{1}{M_m}\begin{vmatrix} a_1^t & b_1^t & c_1^l & c_1^t \\ a_2^t & b_2^t & c_2^l & c_2^t \\ a_3^t & b_3^t & c_3^l & c_3^t \\ a_4^t & b_4^t & c_4^l & c_4^t \end{vmatrix},$$

$$D_m^{tl}(s,0) = \frac{1}{M_m}\begin{vmatrix} b_1^l & a_1^l & c_1^l & c_1^t \\ b_2^l & a_2^l & c_2^l & c_2^t \\ b_3^l & a_3^l & c_3^l & c_3^t \\ b_4^l & a_4^l & c_4^l & c_4^t \end{vmatrix}, \quad D_m^{tt}(s,0) = \frac{1}{M_m}\begin{vmatrix} b_1^l & a_1^t & c_1^l & c_1^t \\ b_2^l & a_2^t & c_2^l & c_2^t \\ b_3^l & a_3^t & c_3^l & c_3^t \\ b_4^l & a_4^t & c_4^l & c_4^t \end{vmatrix},$$

$$\mathbf{M}_m = \begin{vmatrix} b_1^l & b_1^t & c_1^l & c_1^t \\ b_2^l & b_2^t & c_2^l & c_2^t \\ b_3^l & b_3^t & c_3^l & c_3^t \\ b_4^l & b_4^t & c_4^l & c_4^t \end{vmatrix}.$$

where $a_1^l = -k_{l0}J_m'(k_{l0}r_s)$, $a_2^l = \dfrac{im}{r_s}J_m(k_{l0}r_s)$,

$$a_3^l = -(\lambda_0 + 2\mu_0)k_{l0}^2 J_m''(k_{l0}r_s) - \frac{\lambda_0}{r_s}k_{l0}J_m'(k_{l0}r_s) + m^2\frac{\lambda_0}{r_s^2}J_m(k_{l0}r_s),$$

$$a_4^l = -2\frac{im}{r_s}\mu_0 k_{l0}J_m'(k_{l0}r_s) + 2\frac{im}{r_s^2}\mu_0 J_m(k_{l0}r_s), \quad a_1^t = -\frac{im}{r_s}J_m(k_{t0}r_s), \quad a_2^t = -k_{ts}J_m'(k_{t0}r_s),$$

$$a_3^t = 2\frac{im}{r_s^2}\mu_0 J_m(k_{t0}r_s) - 2\frac{im}{r_s}\mu_0 k_{t0}J_m'(k_{t0}r_s),$$

$$a_4^t = \mu_0 k_{t0}^2 J_m''(k_{t0}r_s) - \frac{\mu_0}{r_s}k_{t0}J_m'(k_{t0}r_s) + \mu_0 \frac{m^2}{r_s^2}J_m(k_{t0}r_s), \quad b_1^l = k_{l0}H_m^{(1)'}(k_{l0}r_s),$$

$$b_2^l = -\frac{im}{r_s}H_m(k_{l0}r_s),$$

$$b_3^l = (\lambda_0 + 2\mu_0)k_{l0}^2 H_m^{(1)''}(k_{l0}r_s) + \frac{\lambda_0}{r_s}k_{l0}H_m^{(1)'}(k_{l0}r_s) - m^2\frac{\lambda_0}{r_s^2}H_m^{(1)}(k_{l0}r_s),$$

$$b_4^l = 2\frac{im}{r_s}\mu_0 k_{l0}H_m^{(1)'}(k_{l0}r_s) - 2\frac{im}{r_s^2}\mu_0 H_m^{(1)}(k_{l0}r_s), \quad b_1^t = \frac{im}{r_s}H_m^{(1)}(k_{t0}r_s),$$

$$b_2^t = k_{t0}H_m^{(1)'}(k_{t0}r_s), \quad b_3^t = -2\frac{im}{r_s^2}\mu_0 H_m^{(1)}(k_{t0}r_s) + 2\frac{im}{r_s}\mu_0 k_{t0}H_m^{(1)'}(k_{t0}r_s),$$

$$b_4^t = -\mu_0 k_{t0}^2 H_m^{(1)''}(k_{t0}r_s) + \frac{\mu_0}{r_s}k_{t0}H_m^{(1)'}(k_{t0}r_s) - \mu_0\frac{m^2}{r_s^2}H_m^{(1)}(k_{t0}r_s), \quad c_1^l = -k_{ls}J_m'(k_{ls}r_s),$$

$$c_2^l = \frac{im}{r_s}J_m(k_{ls}r_s), \quad c_3^l = -(\lambda_s + 2\mu_s)k_{ls}^2 J_m''(k_{ls}r_s) - \frac{\lambda_s}{r_s}k_{ls}J_m'(k_{ls}r_s) + m^2\frac{\lambda_s}{r_s^2}J_m(k_{ls}r_s),$$

$$c_4^l = -2\frac{im}{r_s}\mu_s k_{ls}J_m'(k_{ls}r_s) + 2\frac{im}{r_s^2}\mu_s J_m(k_{ls}r_s), \quad c_1^t = -\frac{im}{r_s}J_m(k_{ts}r_s), \quad c_2^t = -k_{ts}J_m'(k_{ts}r_s),$$

$$c_3^t = 2\frac{im}{r_s^2}\mu_s J_m(k_{ts}r_s) - 2\frac{im}{r_s}\mu_s k_{ts}J_m'(k_{ts}r_s),$$

$$c_4^t = \mu_s k_{ts}^2 J_m''(k_{ts}r_s) - \frac{\mu_s}{r_s}k_{ts}J_m'(k_{ts}r_s) + \mu_s\frac{m^2}{r_s^2}J_m(k_{ts}r_s).$$

**Appendix B**

Taking $J_0(x) \cong 1 - \frac{x^2}{4}$ and $H_0^{(1)}(x) \cong i\frac{2}{\pi}\ln x$, for $x \equiv k_{le}r_0$, $k_{te}r_0$, $k_{l0}r_0$ and $k_{t0}r_0$

in $D_m^{\alpha\beta}(e,0)$ when m = 0, we find, to the lowest order in $\omega$:

$$D_0^{ll}(e,0) \cong \frac{i\pi r_0^2 \omega^2 \rho_0(\lambda_0 + \mu_0 - \lambda_e - \mu_e)}{4(\lambda_0 + 2\mu_0)(\mu_0 + \lambda_e + \mu_e)} = \frac{i\pi r_0^2 k_{l0}^2(\kappa_0 - \kappa_e)}{4(\kappa_e + \mu_0)}, \tag{B1-a}$$

$$D_0^{lt}(e,0), D_0^{tl}(e,0), D_0^{tt}(e,0) \cong 0. \tag{B1-b}$$

We take the same approximations in $J_0(x)$ and $H_0^{(1)}(x)$ as above for $x \equiv k_{l0}r_s$

and $k_{t0}r_s$ in $D_0^{\alpha\beta}(s,0)$ to obtain:

$$D_0^{ll}(s,0) \cong \tilde{D}_0^{ll}(s,0), \tag{B2-a}$$

$$D_0^{lt}(s,0), D_0^{tl}(s,0), D_0^{tt}(s,0) \cong 0. \tag{B2-b}$$

Eq.(B-1) together with Eq. (B-2) and the effective medium condition, i.e.

$D_0^{\alpha\beta}(e,0) = D_0^{\alpha\beta}(s,0)$ ($\alpha, \beta = l, t$), gives Eq.(3a).

Also, when m = 1, the Bessel and Hankel functions in $D_1^{\alpha\beta}(e,0)$ take the forms of

$$J_1(x) \cong \frac{x}{2} - \frac{x^3}{16}, H_1^{(1)}(x) \cong i\left(\frac{x}{\pi}\ln\left(\frac{x}{2}\right) - \frac{2}{\pi x}\right), \text{ for } x \equiv k_{le}r_0, k_{te}r_0, k_{l0}r_0 \text{ and } k_{t0}r_0.$$

Then, $D_1^{\alpha\beta}(e,0)$ change into:

$$D_1^{ll}(e,0) \cong -\frac{i\pi r_0^2 \omega^2 (\rho_0 - \rho_e)}{8(\lambda_0 + 2\mu_0)} = -\frac{i\pi r_0^2 k_{l0}^2 (\rho_0 - \rho_e)}{8\rho_0}, \tag{B3-a}$$

$$D_1^{lt}(e,0) = -D_1^{tl}(e,0) \cong \frac{k_{t0}}{k_{l0}} \frac{\pi r_0^2 \omega^2 (\rho_0 - \rho_e)}{8(\lambda_0 + 2\mu_0)}, \tag{B3-b}$$

$$D_1^{tt}(e,0) \cong -\left(\frac{k_{t0}}{k_{l0}}\right)^2 \frac{i\pi r_0^2 \omega^2 (\rho_0 - \rho_e)}{8(\lambda_0 + 2\mu_0)}. \tag{B3-c}$$

The same approximation of $J_1(x)$ and $H_1^{(0)}(x)$ can be taken for $x \equiv k_{l0}r_s$ and $k_{t0}r_s$ in $D_1^{\alpha\beta}(s,0)$. Then $D_1^{\alpha\beta}(s,0)$ change into the following forms:

$$D_1^{ll}(s,0) \cong \tilde{D}_1^{ll}(s,0), \tag{B4-a}$$

$$D_1^{lt}(s,0) = -D_1^{tl}(s,0) \cong i\frac{k_{t0}}{k_{l0}} \tilde{D}_1^{ll}(s,0), \tag{B4-b}$$

$$D_1^{tt}(s,0) \cong \left(\frac{k_{t0}}{k_{l0}}\right)^2 \tilde{D}_1^{ll}(s,0). \tag{B4-c}$$

It is clear to see that $D_1^{lt}(e,0)$, $D_1^{tl}(e,0)$ and $D_1^{tt}(e,0)$ are proportional to $D_1^{ll}(e,0)$ in the same ways as $D_1^{lt}(s,0)$, $D_1^{tl}(s,0)$ and $D_1^{tt}(s,0)$ to $D_1^{ll}(s,0)$. Thus, the four effective medium conditions give four identical equations, which have the form shown in Eq.(3b).

Similarly, when m = 2, we have $J_2(x) \cong \dfrac{x^2}{8} - \dfrac{x^4}{96}$, $H_2^{(1)}(x) \cong i\left(-\dfrac{4}{\pi x^2} - \dfrac{1}{\pi}\right)$ for

$x \equiv k_{le}r_0$, $k_{te}r_0$, $k_{l0}r_0$ and $k_{t0}r_0$ in $D_2^{\alpha\beta}(e,0)$ and obtain

$$D_2^{ll}(e,0) \cong \frac{i\pi r_0^2 \omega^2 \mu_0 \rho_0(\mu_0 - \mu_e)}{4(\lambda_0 + 2\mu_0)(\kappa_0\mu_0 + (\kappa_0 + 2\mu_0)\mu_e)} = \frac{i\pi r_0^2 k_{l0}^2 \mu_0(\mu_0 - \mu_e)}{4(\kappa_0\mu_0 + (\kappa_0 + 2\mu_0)\mu_e)}, \quad \text{(B5-a)}$$

$$D_2^{lt}(e,0) = -D_2^{tl}(e,0) \cong -\left(\frac{k_{t0}}{k_{l0}}\right)^2 \frac{\pi r_0^2 \omega^2 \mu_0 \rho_0(\mu_0 - \mu_e)}{4(\lambda_0 + 2\mu_0)(\kappa_0\mu_0 + (\kappa_0 + 2\mu_0)\mu_e)}, \quad \text{(B5-b)}$$

$$D_2^{tt}(e,0) \cong \left(\frac{k_{t0}}{k_{l0}}\right)^4 \frac{i\pi r_0^2 \omega^2 \mu_0 \rho_0(\mu_0 - \mu_e)}{4(\lambda_0 + 2\mu_0)(\kappa_0\mu_0 + (\kappa_0 + 2\mu_0)\mu_e)}. \quad \text{(B5-c)}$$

Again, after taking the same approximation of $J_2(x)$ and $H_2^{(1)}(x)$ for $x \equiv k_{l0}r_s$ and $k_{t0}r_s$ in $D_2^{\alpha\beta}(s,0)$, we get the following expressions for $D_2^{\alpha\beta}(s,0)$:

$$D_2^{ll}(s,0) \cong \tilde{D}_2^{ll}(s,0), \quad \text{(B6-a)}$$

$$D_2^{lt}(s,0) = -D_2^{tl}(s,0) \cong i\left(\frac{k_{t0}}{k_{l0}}\right)^2 \tilde{D}_2^{ll}(s,0), \quad \text{(B6-b)}$$

$$D_2^{tt}(s,0) \cong \left(\frac{k_{t0}}{k_{l0}}\right)^4 \tilde{D}_2^{ll}(s,0). \quad \text{(B6-c)}$$

We find again, $D_2^{lt}(e,0)$, $D_2^{tl}(e,0)$ and $D_2^{tt}(e,0)$ are proportional to $D_2^{ll}(e,0)$ with the same ratios as those appeared in the ratio of $D_2^{lt}(s,0)$, $D_2^{tl}(s,0)$ and $D_2^{tt}(s,0)$ to $D_2^{ll}(s,0)$, respectively. Then, the last effective medium equation, i.e. Eq.(3c) can be obtained.

**References:**

[1] D. R. Smith W. J. Padilla, D. C. Vier, S. C. Nemat-Nasser, and S. Schultz, Phys. Rev. Lett. **84**, 4184 (2000); R. A. Shelby D. R. Smith, S. C. Nemat-Nasser, and S. Schultz, Appl. Phys. Lett. **78**, 489 (2001); R. A. Shelby, D. R. Smith, and S. Schultz, Science **292**, 77 (2001).

[2] J. B. Pendry, Phys. Rev. Lett. **85**, 3966 (2000); N.Garcia and M.Nieto-Vesperinas, Phys. Rev. Lett. **88,** 207403 (2002)

[3] J. B. Pendry, A. J. Holden, D. J. Robbins, and W. J. Stewart, IEEE Trans.Microwave Theory Tech. **47**, 2075 (1999); J. B. Pendry, A. J. Holden, W. J. Stewart, and I. Youngs, Phys. Rev. Lett. **76**, 4773 (1996).

[4] L. Lewin, Proc. Inst. Elec. Eng., **94**, 65 (1947); A.N. Lagarkov, A.K. Sarychev, Y.R. Smychkovich, and A.P. Vinogradov, J.Electromagn.Waves Appl. **6**, 1159 (1992) ; A. K. Sarychev, R. C. McPhedran, and V. M. Shalaev, Phys.Rev.B **62**, 8531 (2000).

[5] Jensen Li, and C. T. Chan, Phys. Rev. E **70**, 055602 (2004).

[6] Stephen O'Brien and J. B. Pendry, J.Phys.:Condens.Matter **14**, 4035 (2002).

[7] Th. Koschny, P. Markoš, E. N. Economou, D. R. Smith, D. C. Vier, and C. M. Soukoulis, Phys. Rev. B **71**, 245105 (2005).

[8] Y. Wu, J. Li, Z. Q. Zhang, and C. T. Chan, Phys. Rev. B **74**, 085111 (2006).

[9] J. Mei, Z. Liu, W. Wen, and P. Sheng, Phys. Rev. Lett. **96**, 024301 (2006).

[10] Z. Liu X. Zhang, Y. Mao, Y. Y. Zhu, Z. Yang, C. T. Chan, P. Sheng, Science **289**, 1734 (2000) ; C. Goffaux, J. Sánchez-Dehesa, A. Levy Yeyati, Ph. Lambin, A. Khelif, J. O. Vasseur and B. Djafari-Rouhani, Phys. Rev. Lett. **88**, 225502 (2002).

[11] Z. Liu, C. T. Chan, and P. Sheng, Phys. Rev. B **71**, 014103 (2005).

[12] M. S. Kushwaha, P. Halevi, G. Martínez, L. Dobrzynski, and B. Djafari-Rouhani,, Phys. Rev. B **49**, 2313 (1994).

[13] Y. Lai's PhD dissertation: http://lbxml.ust.hk/th_imgo/b909302.pdf (Hong Kong University of Science and Technology, 2005).

[14] M. Kafesaki, R. S. Penciu, and E. N. Economou, Phys. Rev. Lett. **84**, 6050 (2000).

[15] P. Sheng, *Introduction to Wave Scattering, Localization, and Mesoscopic Phenomena* (Academic Press, San Diego, 1995)

[16] W. Lamb, D. M. Wood, and N. W. Ashcroft, Phys. Rev. B **21**, 2248 (1980).

[17] The silicon rubber considered has $\rho = 1.3 \times 10^3 \, kg/m^3$, $\lambda = 6 \times 10^5 \, N/m^2$ and $\mu = 4 \times 10^4 \, N/m^2$. The corresponding parameters in the epoxy host are $\rho = 1.18 \times 10^3 \, kg/m^3$, $\lambda = 4.43 \times 10^9 \, N/m^2$ and $\mu = 1.59 \times 10^9 \, N/m^2$. For lead, $\rho = 11.6 \times 10^3 \, kg/m^3$, $\lambda = 4.23 \times 10^{10} \, N/m^2$ and $\mu = 1.49 \times 10^{10} \, N/m^2$. For air, $\rho = 1.23 \, kg/m^3, k = \lambda = 1.42 \times 10^{10} \, N/m^2$. The above parameters are obtained from Ref.[10].

[18] M. Kafesaki and E. N. Economou, Phys. Rev. B **60**, 11993 (1999); J. Mei, Z. Liu, J. Shi, and D. Tian, Phys. Rev. B **67**, 245107 (2003).

[19] Q. Ni, and J. Cheng, Phys. Rev. B **72**, 014305 (2005)

[20] Y. Wu, Y. Lai, and Z. Q. Zhang (unpublished)

|  | $\kappa_e + \mu_e > 0$ $\mu_e > 0$ | $\kappa_e + \mu_e > 0$ $\mu_e < 0$ | $\kappa_e + \mu_e < 0$ $\mu_e > 0$ | $\kappa_e + \mu_e < 0$ $\mu_e < 0$ |
|---|---|---|---|---|
| $\rho_e > 0$ | $n_l > 0; n_t > 0$ | $n_l > 0; t\text{:}gap$ | $n_t > 0; l\text{:}gap$ | $l\&t\text{: }gap$ |
| $\rho_e < 0$ | $l\&t\text{: }gap$ | $n_t < 0; l\text{:}gap$ | $n_l < 0; t\text{:}gap$ | $n_l < 0; n_t < 0$ |

**Table 1　Various wave propagation properties under different combinations of signs in $\rho_e, \mu_e$ and $\kappa_e + \mu_e$. Positive (negative) *n* indicates positive (negative) propagating bands. *l* and *t* represent longitudinal and transverse waves, respectively.**

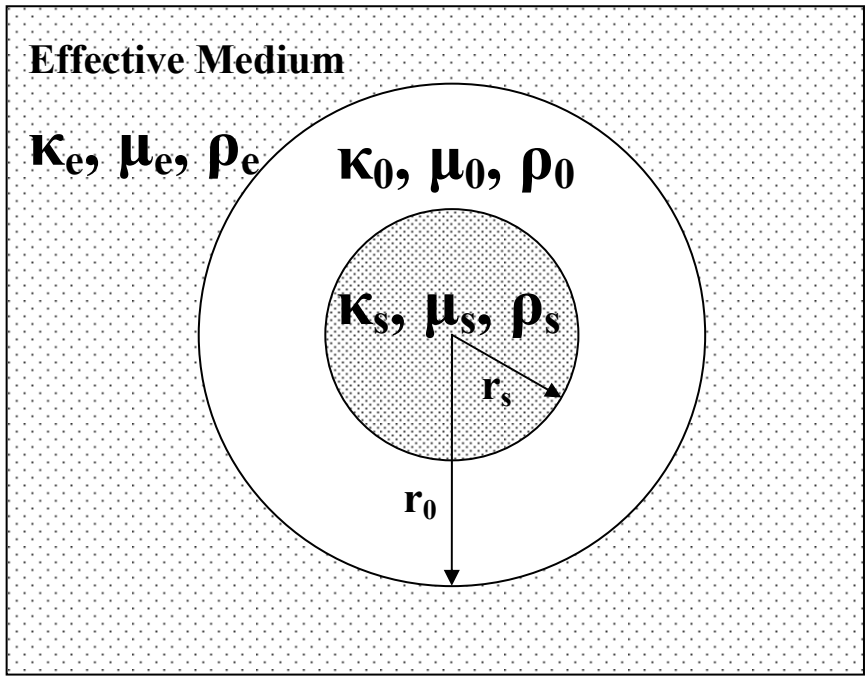

**FIG 1. Microstructure of the effective medium**

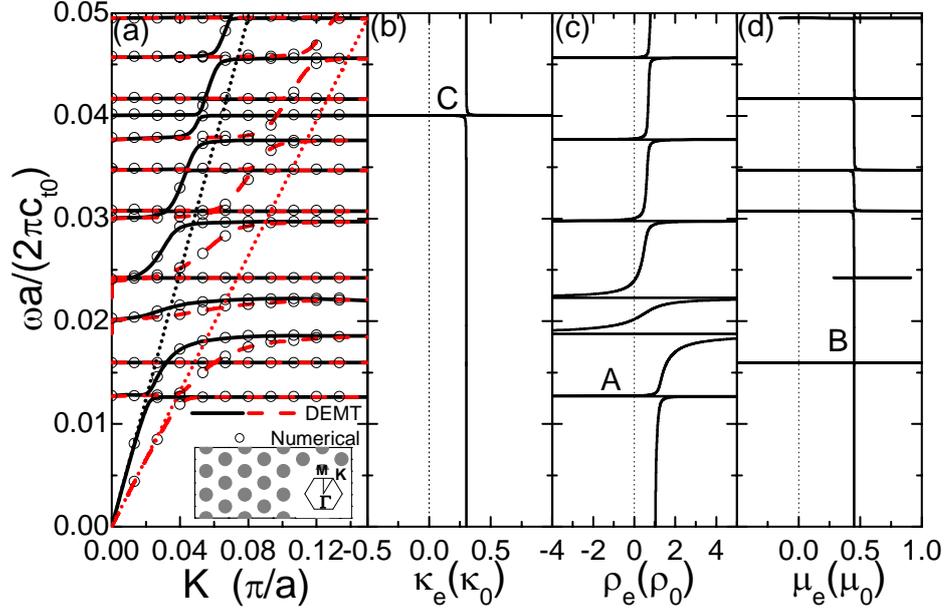

**FIG. 2 (Color online)** (a) Dispersion relation along the $\Gamma K$ direction for a triangular lattice of silicon rubber cylinders ($r_s = 0.3a$) embedded in epoxy host. Open circles are obtained by band-structure calculation and curves are determined from Eq. (3), solid black and dashed red curves represent longitudinal and transverse branches, respectively. Dotted lines are obtained from Eq, (4). (b), (c) and (d) are the effective bulk modulus, effective mass density and effective shear modulus obtained from Eq. (3), respectively.

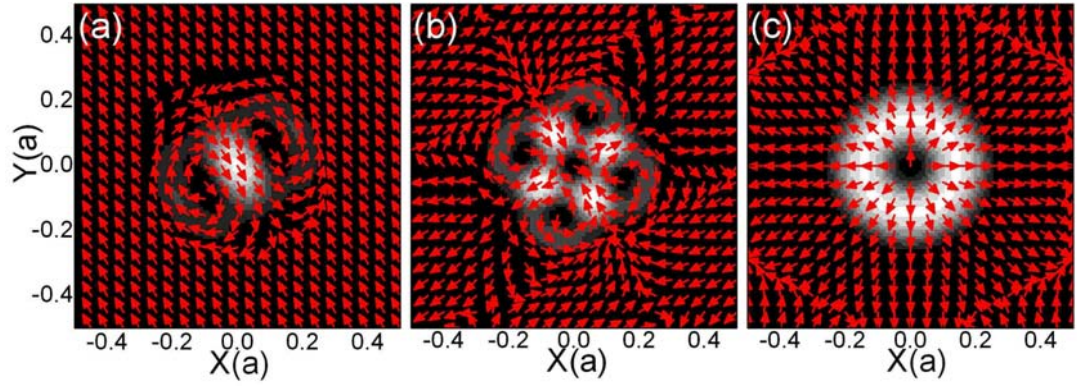

FIG. 3 (Color online) (a), (b) and (c) correspond to the displacement fields at frequencies as marked by "A", "B" and "C" in Fig. 1, respectively. Brightness represents the magnitude and arrows denote the directions.

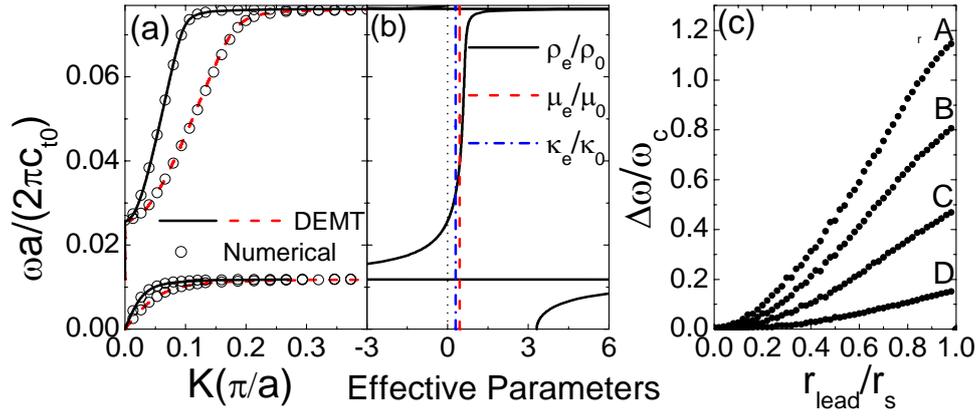

**FIG. 4 (Color online) (a)** The same as in Fig. 1(a) but the central region of the silicon rubber cylinder is replaced by lead cylinder with radius $r_{lead}=0.9r_s$. **(b)** The effective parameters, i.e. $\rho_e$ (solid and black), $\mu_e$ (dashed and red) and $\kappa_e$ (dash-dotted and blue), of the above system obtained from Eq. (3). **(c)** Gap-midgap ratio obtained from Eq. (3). "A", "B", "C" and "D" correspond to $r_s=0.4a$, $0.3a$, $0.2a$ and $0.1a$, respectively.

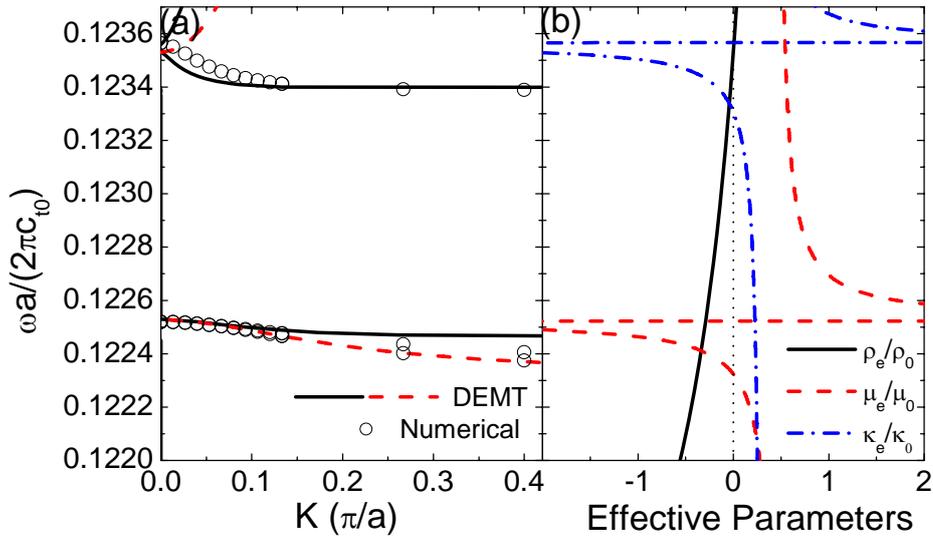

**FIG.5 (Color Online) (a)** The same as in Fig. 1(a) but the central region of the silicon rubber cylinder is replaced by air cylinder with radius $r_{air} = 0.87 r_s$. **(b)** The effective parameters, i.e. $\rho_e$ (solid and black), $\mu_e$ (dashed and red) and $\kappa_e$ (dash-dotted and blue), of the above system obtained from Eq.(3).